\documentclass[fleqn,usenatbib]{mnras}
\usepackage{graphicx}
\usepackage{natbib}
\usepackage{txfonts}
\usepackage{lastpage}
\usepackage{graphics,wrapfig,times}
\usepackage[absolute]{textpos}

\begin{document}

\title[Large fraction of blends from \cite{2023MNRAS.520.2386R} ]{Large fraction of blends from \cite{2023MNRAS.520.2386R}}

 \author[Zasche et al.]{Zasche,~P.\\
  Charles University, Faculty of Mathematics and Physics, Astronomical Institute, V Hole\v{s}ovi\v{c}k\'ach 2, Praha 8, 180 00, Czech Republic, \\ \hspace{0.5cm} e-mail: zasche@sirrah.troja.mff.cuni.cz \\
 }

% \date{Received \today; accepted ???}

\date{\today}
\pagerange{\pageref{firstpage}--\pageref{LastPage}} \pubyear{2025} \maketitle \label{firstpage}

\begin{abstract}
Following the work by Rowan et al. (2023), I found that a large fraction of systems presented as
triple/quadruple stars are in fact only blends. After careful inspection of 225 candidate systems
from Rowan et al. (2023), I found that one third of them are, in fact, only blended signals of two
different eclipsing binaries, well-separated from each other on the sky. Their typical angular
separation is dozens of arcsec, reaching up to more than one hundred arcsec for some targets. This
is probably due to relatively large TESS pixels and insufficient cross-checking of all available
catalogs and photometric data. I present a list of 76 confirmed blends with their correct
identification and respective periods. Such a high fraction (33.8\%) of reported false-positive
systems should be taken into account in future studies, when reporting to new discoveries of
doubly eclipsing candidates.
 \medskip
\end{abstract}


%\maketitle

\section{Introduction}

Recently, \cite{2023MNRAS.520.2386R} identified a number of eclipsing systems showing interesting
photometric behaviour due to what the authors referred to as 'extra-physics'. Among these systems,
the authors presented 225 targets for which two different eclipsing periods are seen in the
lightcurve. The authors labelled these as candidates for triple/quadruple stellar multiples.
However, it is not uncommon for two unrelated eclipsing binaries to contaminate each other's
lightcurve, thus mimicking a genuine doubly-eclipsing quadruple system. The TESS data is
particular prone to such issues due to the large pixels (21$^{\prime\prime}$). Therefore, there is
a considerable risk of incorrect identification of the true source of photometric variability in a
target's lightcurve. This is especially problematic in dense stellar fields. Due to this reason,
we performed a detailed cross-checking of the \cite{2023MNRAS.520.2386R} doubly-eclipsing
candidates and tried to confirm the source of each periodic signal.

This is similar to the procedure we performed for the recent list of new doubly eclipsing
candidates from OGLE \citep{2023A&A...674A.170A}, where we found a false positive rate of about
1/4 \citep{2024A&A...688A..41Z}.

The whole manuscript is organized as follows. Section 2 presents the methods used for the
detection, and Section 3 summarises our results, while Section 4 gives a brief conclusion.

\section{Methods}

Our method of checking the true identification of photometric source was relatively
straightforward. I investigated a 200$^{\prime\prime}$-radius area centered on each
doubly-eclipsing candidate from \cite{2023MNRAS.520.2386R} to detect resolved nearby variable
sources, and measure the corresponding period. For this purpose mostly the VSX catalogue was used
\citep{2006SASS...25...47W}, accompanied with the GAIA DR3 \citep{2023A&A...674A...1G}, ASAS-SN
survey data \citep{2014ApJ...788...48S,2017PASP..129j4502K} , ZTF survey
\citep{2019PASP..131a8003M}, ATLAS \citep{2018AJ....156..241H}, and ASAS
\citep{1997AcA....47..467P}.

Typically several variable stars were found in the 200$^{\prime\prime}$ vicinity of the main
target (but mostly much fainter and of irregular type, or having very long periods of hundreds of
days). However, the correct identification of the source of photometric variability was done due
to its higher brightness (closer to the brightness of the main target), and the proper period
close to the second period given by \cite{2023MNRAS.520.2386R}. In several cases the given period
for a particular target is half or double the value given by \cite{2023MNRAS.520.2386R}, this
information is also given in Table \ref{Tab}.

\section{Results}

From the complete list of the 225 sources, I confirmed 76 blends and identified the corresponding
source of photometric variability. These systems are given in Table \ref{Tab}. The first column
gives the identification of the particular star with both periods, as taken from
\cite{2023MNRAS.520.2386R}. In their work, \cite{2023MNRAS.520.2386R} also presented their Table
2, where they listed the systems close to the main target noted as variables as well. However, we
found only 13 such systems among our detected 76, i.e., only a small fraction. The rest of the
systems presented are our confirmed blends for the first time here.

\begin{table*}
  \caption{False positive systems from Rowan et al. (2023)  %\cite{2023MNRAS.520.2386R}
  together with a correct source identification.}
  \label{Tab}
%  \centering
  \scalebox{0.84}{
  \begin{tabular}{c c c c c c c c}\\[-3mm]
\hline \hline\\[-3mm]
  Information from            &  Published/catalog  &  Identification & TESS number &     RA     &     DE     & Remark  & Angular distance  \\
  \cite{2023MNRAS.520.2386R}  &  period   &              &   &  [J2000.0] & [J2000.0]  &         &   \\
 \hline
 ASASSN-V J022611.76+391422.2  &   &    &   &  &  &  1 &    \\
 = TIC 129913429  &   &    &   &  &  &    &    \\
 5.44807446 d   &  5.4477  d  & ASASSN-V J022611.72+391423.8 & TIC 129913429 & 02 26 11.72 & +39 14 23.82 &    &  8$^{\prime\prime}$ \\
 14.0894846 d   & 6.83606525 d&  ZTF J022611.88+391415.8   & TIC 129913431  & 02 26 11.89 & +39 14 15.94 & half period &    \\
  \hline
   ASASSN-V J051003.14+383555.1  &   &    &   &  &  &   &    \\
   = TIC 122878320  &   &    &   &  &  &   &    \\
  2.65425561 d & 2.6541552 d  & ASASSN-V J051003.14+383555.1 & TIC 122878320 & 05 10 03.15 & +38 35 55.21 &   & 67$^{\prime\prime}$ \\
  1.1097072  d & 1.1103227 d  & ASASSN-V J051003.09+383447.1 & TIC 122878283 & 05 10 03.07 & +38 34 46.56 &   &  \\
  \hline
  ASASSN-V J051232.84+481910.3  &   &    &   &  &   &   &    \\
  = TIC 355499690  &   &    &   &  &   &   &    \\
  1.72216671 d & 1.7219936 d &  ZTF J051232.83+481910.7     & TIC 355499690 & 05 12 32.83 & +48 19 10.70 &   & 43$^{\prime\prime}$ \\
  3.6476492  d & 7.2853168 d & Gaia DR3 213185683902601344  & TIC 355499680 & 05 12 28.73 & +48 19 22.76 & double period  & \\
  \hline
  ASASSN-V J052123.95+413756.6  &   &  &  &   &   &   &    \\
  = TIC 426671266  &   &  &  &   &   &   &    \\
  5.74182481 d & 5.7431314 d & ZTF J052123.93+413756.3      & TIC 426671266 & 05 21 23.94 & +41 37 56.32 &    & 25$^{\prime\prime}$ \\
  1.6531282  d & 1.6523632 d & ASASSN-V J052126.07+413749.8 & TIC 426671263 & 05 21 26.10 & +41 37 50.99 &    & \\
  \hline
   ASASSN-V J053118.97+373717.6  &   &  &  &   &   &   &    \\
   = TIC 67026938  &   &  &  &   &   &   &    \\
  3.34657868 d & 3.3465993 d & ASASSN-V J053119.02+373718.3 & TIC 67026938 & 05 31 18.98 & +37 37 17.62 &    & 74$^{\prime\prime}$ \\
  5.0396235  d & 5.025350  d & ASASSN-V J053114.42+373809.9 & TIC 67026910 & 05 31 14.50 & +37 38 08.99 &    & \\
  \hline
   ASASSN-V J055645.83-731943.4  &   &  &  &   &   &   &    \\
   = TIC 141622079 &   &  &  &   &   &   &    \\
  5.48504770 d & 5.484878 d & WISE J055645.8-731943         & TIC 141622079 & 05 56 45.84 & -73 19 43.21 &    & 64$^{\prime\prime}$ \\
  2.9279463  d & 2.92794  d & ASAS J055655-7320.5           & TIC 141622065 & 05 56 54.90 & -73 20 33.50 &    & \\
  \hline
   ASASSN-V J062731.53-015356.4  &   &  &  &   &   &   &    \\
   = TIC 42766545  &   &  &  &   &   &   &    \\
  2.82041377 d & 2.8207396 d & ASASSN-V J062731.51-015356.2 & TIC 42766545 & 06 27 31.51 & -01 53 56.15 &    & 27$^{\prime\prime}$ \\
  1.5021422  d & 1.5021308 d &  ZTF J062732.54-015333.4     & TIC 42766559 & 06 27 32.54 & -01 53 33.47 &    & \\
  \hline
   ASASSN-V J063034.38-013822.1  &   &  &  &   &   &   &    \\
   = TIC 43250275  &   &  &  &   &   &   &    \\
  4.10498412 d & 2.052508 d  &  ASAS J063034-0138.4         & TIC 43250275 & 06 30 34.37 & -01 38 22.09 & half period & 14$^{\prime\prime}$ \\
  0.9570587  d & 0.95706 d   &  VSSP J063034.86-013833.6    & TIC 43250266 & 06 30 34.86 & -01 38 33.61 &    & \\
  \hline
   ASASSN-V J064605.45+122320.3  &   &  &  &   &   &   &    \\
   = TIC 372490054  &   &  &  &   &   &   &    \\
  7.24902207 d & 7.2481974 d & ASASSN-V J064605.45+122320.3 & TIC 372490054 & 06 46 05.52 & +12 23 19.82 &    & 94$^{\prime\prime}$ \\
  2.0139013  d & 2.01391 d   &  ASAS J064606+1224.8         & TIC 372490191 & 06 46 06.04 & +12 24 53.78 &    & \\
  \hline
   ASASSN-V J065947.05-105704.4  &   &  &  &   &   &   &    \\
   = TIC 124943625  &   &  &  &   &   &   &    \\
  2.62041293 d & 1.3102077 d & ASASSN-V J065947.10-105704.8 & TIC 124943625 & 06 59 47.06 & -10 57 04.54 & half period & 43$^{\prime\prime}$ \\
  0.7016615  d & 0.7016888 d &  GDS$\_$J0659474-105621      & TIC 124943596 & 06 59 47.47 & -10 56 21.41 &    & \\
   \hline
   ASASSN-V J070548.94-133641.5  &   &  &  &   &   &   &    \\
   = TIC 148500081  &   &  &  &   &   &   &    \\
  3.57672855 d & 3.5767469 d & Gaia DR3 3044725856858853120 & TIC 148500081 & 07 05 48.95 & -13 36 41.71 &    & 80$^{\prime\prime}$ \\
  1.3254868  d & 1.3254611 d & Gaia DR3 3044725989999551872 & TIC 148500107 & 07 05 53.47 & -13 35 56.01 &    & \\
  \hline
   ASASSN-V J071438.17-250324.0  &   &  &  &   &   &   &    \\
   = TIC 65815087  &   &  &  &   &   &   &    \\
  3.84402955 d & 3.8438807 d & Gaia DR3 5617183427739613440 & TIC 65815087 & 07 14 38.21 & -25 03 23.30 &    & 43$^{\prime\prime}$ \\
  1.3549545  d & 1.3548761 d & Gaia DR3 5617183423434072704 & TIC 65815077 & 07 14 41.05 & -25 03 04.30 &    & \\
  \hline
   ASASSN-V J071459.48-145101.7  &   &  &  &   &   &   &    \\
    = TIC 295403854  &   &  &  &   &   &   &    \\
  15.1126971 d & 15.111547 d & GDS$\_$J0714595-145101       & TIC 295403854 & 07 14 59.52 & -14 51 01.33 &    & 82$^{\prime\prime}$ \\
   2.3162209 d & 2.3162237 d & Gaia DR3 3032208393712100608 & TIC 295617233 & 07 15 04.51 & -14 50 23.53 &    & \\
   \hline
   ASASSN-V J071505.57+081534.4  &   &  &  &   &   &   &    \\
   = TIC 264313741  &   &  &  &   &   &   &    \\
  1.73758657 d & 1.7376414 d & ASASSN-V J071505.55+081533.1 & TIC 264313741 & 07 15 05.55 & +08 15 33.05 &    & 23$^{\prime\prime}$ \\
  0.6806757  d & 0.6813544 d &   ZTF J071506.91+081543.7    & TIC 264313738 & 07 15 06.92 & +08 15 43.70 &    &  \\
  \hline
   ASASSN-V J071828.98-201859.5  &   &  &  &   &   &   &    \\
   = TIC 4783276 &   &  &  &   &   &   &    \\
  1.08446578 d & 1.0844837 d & ASASSN-V J071828.98-201859.9 & TIC 4783276 & 07 18 28.98 & -20 18 59.66 &    & 27$^{\prime\prime}$ \\
  0.6973733  d & 0.6973951 d & ASASSN-V J071830.87-201854.5 & TIC 4783271 & 07 18 30.87 & -20 18 54.13 &    & \\
  \hline
   ASASSN-V J072136.08-053010.7  &   &  &  &   &   &   &    \\
   = TIC 10175811  &   &  &  &   &   &   &    \\
  10.7789181 d & 10.7792 d   &  GDS$\_$J0721368-053010      & TIC 10175811 & 07 21 36.08 & -05 30 10.87 &    & 52$^{\prime\prime}$ \\
  0.8527995  d & 0.852635 d  & ASASSN-V J072139.42-053024.4 & TIC 10175820 & 07 21 39.42 & -05 30 24.34 &    & \\
  \hline  \hline
\end{tabular}}
 {\small Remarks: $1$ - Systems presented in Table 2 of \cite{2023MNRAS.520.2386R}.}
\end{table*}

\begin{table*}
  \caption{False positive systems from Rowan et al. (2023)  %\cite{2023MNRAS.520.2386R}
  together with a correct source identification.}
%  \centering
  \scalebox{0.84}{
  \begin{tabular}{c c c c c c c c}\\[-3mm]
\hline \hline\\[-3mm]
  Information from            &  Published/catalog  &  Identification & TESS number &     RA     &     DE     & Remark  & Angular distance  \\
  \cite{2023MNRAS.520.2386R}  &  period   &              &   &  [J2000.0] & [J2000.0]  &         &   \\
  \hline
   ASASSN-V J072358.28-255425.8  &   &  &  &   &   &   &    \\
   = TIC 107342914  &   &  &  &   &   &   &    \\
  2.90562949 d & 5.8113652 d & ASASSN-V J072358.31-255426.4 & TIC 107342914 & 07 23 58.31 & -25 54 26.35 & double period & 66$^{\prime\prime}$ \\
  1.6703193  d & 1.67197 d   &   ASAS J072400-2553.4        & TIC 107342976 & 07 24 00.00 & -25 53 24.00 &    & \\
  \hline
   ASASSN-V J072528.80-122300.4  &   &  &  &   &   &   &    \\
   = TIC 386035043  &   &  &  &   &   &   &    \\
  6.30301394 d & 6.3039241 d & Gaia DR3 3034416831533156480 & TIC 386035043 & 07 25 28.83 & -12 23 00.41 &    & 44$^{\prime\prime}$ \\
  0.8342015  d & 0.8342060 d & Gaia DR3 3034416938918134656 & TIC 386035033 & 07 25 31.71 & -12 22 46.86 &    & \\
  \hline
   ASASSN-V J072536.81-282136.5  &   &  &  &   &   &   &    \\
   = TIC 107667146  &   &  &  &   &   &   &    \\
  4.71137547 d & 4.7117408 d & ASASSN-V J072536.81-282136.5 & TIC 107667146 & 07 25 36.79 & -28 21 38.18 &    & 43$^{\prime\prime}$ \\
  0.4706076  d & 0.470837  d & Gaia DR3 5611871583990424960 & TIC 107667114 & 07 25 35.41 & -28 20 58.78 &    & \\
  \hline
   ASASSN-V J072925.58-154309.6  &   &  &  &   &   &   &    \\
   = TIC 50281417  &   &  &  &   &   &   &    \\
  1.58272928 d & 1.5826917 d & Gaia DR3 3028436866972401792 & TIC 50281417 & 07 29 25.59 & -15 43 09.62 &    & 45$^{\prime\prime}$ \\
  3.5726602  d & 3.5727591 d & Gaia DR3 3028436970051604224 & TIC 50477999 & 07 29 27.98 & -15 42 40.60 &    & \\
  \hline
   ASASSN-V J073516.89-221146.9  &   &  &  &   &   & 1  &    \\
   = TIC 348740785  &   &  &  &   &   &    &    \\
  5.52923059 d & 5.5287592 d & ASASSN-V J073516.85-221146.5 & TIC 348740785 & 07 35 16.93 & -22 11 47.26 &     & 62$^{\prime\prime}$ \\
  8.3001797  d & 8.3019512 d & ASASSN-V J073521.11-221129.4 & TIC 348740825 & 07 35 21.13 & -22 11 27.67 &     & \\
   \hline
   ASASSN-V J073521.11-221129.4      &   &  &  &   &   & 1 &    \\
   = TIC 348740825     &   &  &  &   &   &   &    \\
  8.30088452 d & 8.3019512 d & ASASSN-V J073521.11-221129.4 & TIC 348740825 & 07 35 21.13 & -22 11 27.67 &     & 62$^{\prime\prime}$ \\
  5.5292700  d & 5.5287592 d & ASASSN-V J073516.85-221146.5 & TIC 348740785 & 07 35 16.93 & -22 11 47.26 &     & \\
  \hline
   ASASSN-V J074535.54-020311.8  &   &  &  &   &   &   &    \\
   = TIC 68199549 &   &  &  &   &   &   &    \\
  2.90614044 d & 2.9064128 d & ASASSN-V J074535.55-020311.8 & TIC 68199549 & 07 45 35.55 & -02 03 11.92 &     & 47$^{\prime\prime}$ \\
  0.3223560  d & 0.3223548 d &  WISE J074532.7-020330       & TIC 68199538 & 07 45 32.71 & -02 03 30.89 &     & \\
  \hline
   ASASSN-V J074626.41-074013.6  &   &  &  &   &   &   &    \\
   = TIC 31940129  &   &  &  &   &   &   &    \\
  4.60117383 d & 4.601352  d & ASASSN-V J074626.41-074013.6 & TIC 31940129 & 07 46 26.39 & -07 40 13.40 &     & 25$^{\prime\prime}$ \\
  0.7067873  d & 0.7067883 d & Gaia DR3 3043417678537105408 & TIC 31940125 & 07 46 24.94 & -07 40 26.76 &     & \\
  \hline
   ASASSN-V J074905.31-004313.1  &   &  &  &   &   &   &    \\
   = TIC 426059799  &   &  &  &   &   &   &    \\
  4.60723447 d & 4.607259 d  &  ASAS J074905-0043.1         & TIC 426059799 & 07 49 05.32 & -00 43 13.40 &     & 92$^{\prime\prime}$ \\
  1.3758202  d & 0.6875198 d & Gaia DR3 3085572885626407424 & TIC 68475288  & 07 49 00.35 & -00 42 19.62 & half period & \\
  \hline
   ASASSN-V J080732.02-304508.0  &   &  &  &   &   &   &    \\
   = TIC 144801963 &   &  &  &   &   &   &    \\
  9.12344999 d & 9.1220892 d & ASASSN-V J080732.03-304509.0 & TIC 144801963 & 08 07 32.06 & -30 45 08.39 &     & 32$^{\prime\prime}$ \\
  1.5313335  d & 1.5313477 d & Gaia DR3 5596375926105940480 & TIC 144801928 & 08 07 32.34 & -30 44 36.16 &     & \\
  \hline
   ASASSN-V J081525.20+102352.5  &   &  &  &   &   &   &    \\
   = TIC 443956777  &   &  &  &   &   &   &    \\
  3.02012410 d & 3.0202992 d & ASASSN-V J081525.20+102352.5 & TIC 443956777 & 08 15 25.20 & +10 23 52.51 &     & 37$^{\prime\prime}$ \\
  0.3962984  d & 0.396292 d  & CSS$\_$J081526.1+102427      & TIC 443956770 & 08 15 26.14 & +10 24 27.22 &     & \\
  \hline
   ASASSN-V J082911.84-410634.9  &   &  &  &   &   &   &    \\
   = TIC 184510069  &   &  &  &   &   &   &    \\
  7.88725165 d & 7.8871452 d & ASASSN-V J082911.81-410634.9 & TIC 184510069 & 08 29 11.81 & -41 06 34.92 &     & 125$^{\prime\prime}$ \\
  3.7619698  d & 1.881522 d  &   HD 72065                   & TIC 184510206 & 08 29 17.84 & -41 08 19.28 & half period & \\
  \hline
   ASASSN-V J083345.31+120746.1  &   &  &  &   &   &   &    \\
   = TIC 20537406  &   &  &  &   &   &   &    \\
  1.17097787 d & 1.1709737 d & CRTS J083345.2+120746        & TIC 20537406  & 08 33 45.27 & +12 07 46.31 &     & 24$^{\prime\prime}$ \\
  1.4401864  d & 0.720093 d  & Gaia DR3 602596319790932864  & TIC 500552207 & 08 33 45.90 & +12 08 13.00 & half period & \\
  \hline
   ASASSN-V J083424.23-232101.3  &   &  &  &   &   &   &    \\
   = TIC 118063447  &   &  &  &   &   &   &    \\
  4.13603265 d & 4.1357821 d & ASASSN-V J083424.24-232100.9 & TIC 118063447 & 08 34 24.24 & -23 21 00.86 &     & 57$^{\prime\prime}$ \\
  1.5909047  d & 1.5908628 d & ASASSN-V J083428.34-232104.1 & TIC 118063449 & 08 34 28.35 & -23 21 04.07 &     & \\
  \hline
   ASASSN-V J084200.53-301323.3  &   &  &  &   &   &   &    \\
   = TIC 185703116  &   &  &  &   &   &   &    \\
  6.88060673 d & 6.8807102 d & ASASSN-V J084200.53-301323.7 & TIC 185703116 & 08 42 00.53 & -30 13 23.70 &     & 47$^{\prime\prime}$ \\
  0.3995645  d & 0.3995592 d &   WISE J084202.0-301240      & TIC 185703146 & 08 42 02.04 & -30 12 40.50 &     & \\
  \hline
   ASASSN-V J084212.00-464015.6  &   &  &  &   &   &   &    \\
   = TIC 285704031  &   &  &  &   &   &   &    \\
  31.6930419 d & 15.8511306 d & GDS$\_$J0842119-464016      & TIC 285704031 & 08 42 11.92 & -46 40 16.10 & half period & 42$^{\prime\prime}$ \\
  1.3328266  d & 1.332811  d & ASASSN-V J084214.05-464051.9 & TIC 285703974 & 08 42 14.04 & -46 40 51.85 &     & \\
  \hline \hline
   \end{tabular}}
 {\small Remarks: $1$ - Systems presented in Table 2 of \cite{2023MNRAS.520.2386R}.}
\end{table*}

\begin{table*}
  \caption{False positive systems from Rowan et al. (2023)  %\cite{2023MNRAS.520.2386R}
  together with a correct source identification.}
%  \centering
  \scalebox{0.84}{
  \begin{tabular}{c c c c c c c c}\\[-3mm]
\hline \hline\\[-3mm]
  Information from            &  Published/catalog  &  Identification & TESS number &     RA     &     DE     & Remark  & Angular distance  \\
  \cite{2023MNRAS.520.2386R}  &  period   &              &   &  [J2000.0] & [J2000.0]  &         &   \\
  \hline
    ASASSN-V J091505.23-421038.2  &   &  &  &   &   &   &    \\
    = TIC 75399656  &   &  &  &   &   &   &    \\
  9.21445852 d & 9.2138838 d & WISE J091505.2-421038        & TIC 75399656 & 09 15 05.23 & -42 10 38.39 &     & 68$^{\prime\prime}$ \\
  3.3983392  d & 3.3982442 d & Gaia DR3 5427697545177221248 & TIC 75147782 & 09 15 00.33 & -42 09 56.88 &     & \\
  \hline
    ASASSN-V J094218.42-511944.2  &   &  &  &   &   &  1 &    \\
    = TIC 362809861  &   &  &  &   &   &    &    \\
  1.63627505 d & 1.6362964 d & ASASSN-V J094218.41-511943.7 & TIC 362809861 & 09 42 18.41 & -51 19 43.75 &     & 19$^{\prime\prime}$ \\
  2.2927765  d & 2.2928046 d & Gaia DR3 5405816439166550912 & TIC 362809839 & 09 42 19.14 & -51 19 26.51 &     & \\
  \hline
   ASASSN-V J094829.09-450911.9  &   &  &  &   &   &   &    \\
   = TIC 869130392 &   &  &  &   &   &   &    \\
  1.44753117 d & 1.4475447 d & ASASSN-V J094829.12-450913.0 & TIC 869130392 & 09 48 29.12 & -45 09 12.99 &     & 29$^{\prime\prime}$ \\
  1.2782598  d & 1.2780882 d & Gaia DR3 5411633233623280128 & TIC 34731735  & 09 48 26.47 & -45 09 20.52 &     & \\
  \hline
   ASASSN-V J103509.23-465456.8  &   &  &  &   &   &   &    \\
   = TIC 146485604  &   &  &  &   &   &   &    \\
  0.94199008 d & 0.942006 d & ASASSN-V J103509.25-465456.9  & TIC 146485604 & 10 35 09.25 & -46 54 56.92 &     & 34$^{\prime\prime}$ \\
  6.1791427  d & 6.179    d & Gaia DR3 5364890692219940736  & TIC 146485621 & 10 35 09.01 & -46 55 30.97 &     & \\
  \hline
   ASASSN-V J103615.81-520120.0  &   &  &  &   &   &   &    \\
   = TIC 303109583  &   &  &  &   &   &   &    \\
  2.42055423 d & 4.8411269 d & ASASSN-V J103615.91-520103.0 & TIC 303109573 & 10 36 15.91 & -52 01 03.04 & double period & 89$^{\prime\prime}$ \\
  0.4221929  d & 0.4221946 d &   ASAS J103617-5202.5        & TIC 303109640 & 10 36 17.34 & -52 02 31.42 &     & \\
  \hline
   ASASSN-V J104141.53-590606.7  &   &  &  &   &   &   &    \\
   = TIC 458567336  &   &  &  &   &   &   &    \\
  3.57252416 d & 3.5724527 d & Gaia DR3 5350596560036697216 & TIC 458567336 & 10 41 41.52 & -59 06 06.58 &     & 39$^{\prime\prime}$ \\
  4.9805081  d & 2.4911785 d & Gaia DR3 5350596491317226368 & TIC 458567335 & 10 41 46.53 & -59 06 06.79 & half period & \\
   \hline
   ASASSN-V J105907.86-620142.6  &   &  &  &   &   &  1 &    \\
   = TIC 465971173  &   &  &  &   &   &    &    \\
  1.26150468 d & 1.2614944 d & GDS$\_$J1059076-620141       & TIC 465971173 & 10 59 07.67 & -62 01 41.30 &     & 4$^{\prime\prime}$ \\
  1.2830471  d & 1.2830361 d & Gaia DR3 5241801705522810240 & TIC 465971173 & 10 59 07.87 & -62 01 37.56 &     & \\
  \hline
   ASASSN-V J111231.92-584035.6  &   &  &  &   &   &   &    \\
   = TIC 309081901  &   &  &  &   &   &   &    \\
  1.94934078 d & 1.9493503 d & ASAS J111232-5840.6          & TIC 309081901 & 11 12 32.00 & -58 40 36.01 &     & 44$^{\prime\prime}$ \\
  7.5517054  d & 3.7772258 d &  GDS$\_$J1112325-583951      & TIC 309081983 & 11 12 32.53 & -58 39 51.88 & half period & \\
  \hline
   ASASSN-V J111730.54-533230.2  &   &  &  &   &   &   &    \\
   = TIC 82488083  &   &  &  &   &   &   &    \\
  5.07227980 d & 5.0703467 d & ASASSN-V J111730.53-533230.7 & TIC 82488083 & 11 17 30.53 & -53 32 30.73 &     & 59$^{\prime\prime}$ \\
  4.5762662  d & 4.5639614 d & ASASSN-V J111736.20-533200.1 & TIC 82488061 & 11 17 36.20 & -53 32 00.13 &     & \\
  \hline
   ASASSN-V J112257.67-614734.9  &   &  &  &   &   &   &    \\
   = TIC 280381105  &   &  &  &   &   &   &    \\
  3.31352572 d & 3.3136889 d &  V0440 Cen                   & TIC 280381105 & 11 22 57.65 & -61 47 35.59 &     & 51$^{\prime\prime}$ \\
  2.6768711  d & 2.676980  d & ASAS J112301-6146.8          & TIC 280380993 & 11 23 01.60 & -61 46 53.00 &     & \\
  \hline
   ASASSN-V J112301.25-614651.5  &   &  &  &   &   &   &    \\
   = TIC 280380993  &   &  &  &   &   &   &    \\
  2.67696477 d & 2.676980  d & ASAS J112301-6146.8          & TIC 280380993 & 11 23 01.60 & -61 46 53.00 &    & 51$^{\prime\prime}$ \\
  3.3135030  d & 3.3136889 d &  V0440 Cen                   & TIC 280381105 & 11 22 57.65 & -61 47 35.59 &    & \\
  \hline
   ASASSN-V J114159.82-614153.0  &   &  &  &   &   &   &    \\
   = TIC 321465152  &   &  &  &   &   &   &    \\
  3.44776265 d & 3.4471023 d & ASASSN-V J114159.76-614156.3 & TIC 321465152 & 11 41 59.76 & -61 41 56.29 &    & 96$^{\prime\prime}$ \\
  19.315147  d & 9.657305 d  &  MO Cen                      & TIC 321464887 & 11 42 01.18 & -61 40 21.29 & half period & \\
  \hline
   ASASSN-V J114316.95-624022.4  &   &  &  &   &   &   &    \\
   = TIC 321946750  &   &  &  &   &   &   &    \\
  3.23895292 d & 3.2389 d    & GDS$\_$J1143169-624022       & TIC 321946750 & 11 43 16.97 & -62 40 22.48 &    & 66$^{\prime\prime}$ \\
  1.2856685  d & 1.285701 d  & ASASSN-V J114318.91-624127.7 & TIC 322420705 & 11 43 18.74 & -62 41 26.99 &    & \\
  \hline
   ASASSN-V J114836.93-625955.3  &   &  &  &   &   &   &    \\
   = TIC 917240260  &   &  &  &   &   &   &    \\
  3.33514512 d & 3.3346516 d & V0787 Cen                    & TIC 917240260 & 11 48 36.73 & -62 59 56.62 &    & 37$^{\prime\prime}$ \\
  0.3031804 d  & 0.6064278 d & Gaia DR3 5333263171560093696 & TIC 324662281 & 11 48 37.86 & -62 59 20.33 & double period & \\
  \hline
   ASASSN-V J115107.59-635508.6  &   &  &  &   &   &   &    \\
   = TIC 304415685  &   &  &  &   &   &   &    \\
  1.63716626 d & 1.6371701 d & Gaia DR3 5332949497176313728 & TIC 304415685 & 11 51 07.64 & -63 55 07.36 &    & 36$^{\prime\prime}$ \\
  3.8290891  d & 3.8290870 d & Gaia DR3 5332949501524752512 & TIC 304415604 & 11 51 06.03 & -63 54 33.26 &    & \\
  \hline
    ASASSN-V J115432.85-610607.2  &   &  &  &   &   &  1 &    \\
    = TIC 305616260  &   &  &  &   &   &    &    \\
  8.72678049 d & 8.7267695 d &  MS Cen                      & TIC 305616260 & 11 54 32.42 & -61 06 09.50 &    & 57$^{\prime\prime}$ \\
  6.3703186  d & 6.3713727 d & ASASSN-V J115430.48-610705.7 & TIC 305616118 & 11 54 30.52 & -61 07 04.37 &    & \\
  \hline\hline
   \end{tabular}}
 {\small Remarks: $1$ - Systems presented in Table 2 of \cite{2023MNRAS.520.2386R}.}
\end{table*}

\begin{table*}
  \caption{False positive systems from Rowan et al. (2023)  %\cite{2023MNRAS.520.2386R}
  together with a correct source identification.}
%  \centering
  \scalebox{0.84}{
  \begin{tabular}{c c c c c c c c}\\[-3mm]
\hline \hline\\[-3mm]
  Information from            &  Published/catalog  &  Identification & TESS number &     RA     &     DE     & Remark  & Angular distance  \\
  \cite{2023MNRAS.520.2386R}  &  period   &              &   &  [J2000.0] & [J2000.0]  &         &   \\
  \hline
   ASASSN-V J122745.12-620656.7  &   &  &  &   &   &   &    \\
   = TIC 450684840  &   &  &  &   &   &   &    \\
  3.34775729 d & 3.3478478 d & Gaia DR3 6054662474357654912 & TIC 450684840 & 12 27 45.18 & -62 06 56.68 &    & 27$^{\prime\prime}$ \\
  2.0372869  d & 2.0373020 d & Gaia DR3 6054662405638174208 & TIC 450684878 & 12 27 48.32 & -62 07 12.42 &    & \\
  \hline
   ASASSN-V J123008.19-640228.0  &   &  &  &   &   &   &    \\
   = TIC 451015749  &   &  &  &   &   &   &    \\
  1.47576235 d & 1.4757595 d & Gaia DR3 6053465174934083456 & TIC 451015749 & 12 30 08.08 & -64 02 26.19 &    & 58$^{\prime\prime}$ \\
  0.5574097  d & 1.1148116 d & ASASSN-V J123001.82-640306.7 & TIC 451015644 & 12 30 01.79 & -64 03 07.02 & double period & \\
  \hline
   ASASSN-V J125309.42-591431.2  &   &  &  &   &   &   &    \\
   = TIC 412633616  &   &  &  &   &   &   &    \\
  2.03393547 d & 2.0338 d    & Gaia DR3 6057084556691224192 & TIC 412633616 & 12 53 09.50 & -59 14 31.70 &    & 26$^{\prime\prime}$ \\
  3.7809746  d & 3.78   d    & Gaia DR3 6057085312605673344 & TIC 412633671 & 12 53 09.17 & -59 14 05.12 &    & \\
  \hline
   ASASSN-V J135105.28-615930.5  &   &  &  &   &   &   &    \\
   = TIC 299182985  &   &  &  &   &   &   &    \\
  2.07336116 d & 2.0734677 d & ASASSN-V J135105.13-615930.6 & TIC 299182985 & 13 51 05.32 & -61 59 30.55 &    & 45$^{\prime\prime}$ \\
  1.4275213  d & 1.4275071 d & Gaia DR3 5865560115171491072 & TIC 299183110 & 13 51 07.97 & -61 58 49.80 &    & \\
  \hline
   ASASSN-V J143028.86-593808.0  &   &  &  &   &   &   &    \\
   = TIC 299182985  &   &  &  &   &   &   &    \\
  3.34503305 d & 3.3457944 d & ASASSN-V J143028.86-593808.0 & TIC 299182985 & 14 30 28.88 & -59 38 07.70 &    & 36$^{\prime\prime}$ \\
  1.1335664  d & 1.1338123 d &   ASAS J143025-5937.9        & TIC 291286216 & 14 30 24.61 & -59 37 51.63 &    & \\
  \hline
   ASASSN-V J145653.00-603725.4  &   &  &  &   &   & 1 &    \\
   = TIC 296815995  &   &  &  &   &   &   &    \\
  5.10103169 d & 5.1012795 d & Gaia DR3 5877942196658162816 & TIC 296815995 & 14 56 52.90 & -60 37 25.77 &    & 50$^{\prime\prime}$ \\
  2.4719043  d & 2.4717569 d & Gaia DR3 5877942261052344960 & TIC 296815901 & 14 56 47.32 & -60 36 57.53 &    & \\
  \hline
   ASASSN-V J160130.92-503017.1  &   &  &  &   &   &   &    \\
   = TIC 275164175  &   &  &  &   &   &   &    \\
  3.30375370 d & 3.3040291 d & ASASSN-V J160130.70-503018.9 & TIC 275164175 & 16 01 30.71 & -50 30 18.90 &    & 52$^{\prime\prime}$ \\
  4.8818492  d & 4.8820967 d & ASASSN-V J160130.98-502926.8 & TIC 275164023 & 16 01 30.98 & -50 29 26.81 &    & \\
  \hline
   ASASSN-V J161358.14-511051.3  &   &  &  &   &   & 1 &    \\
   = TIC 409764822  &   &  &  &   &   &   &    \\
  4.58904638 d & 4.58908 d   & TYC 8323-219-1               & TIC 409764822 & 16 13 58.14 & -51 10 51.38 &    & 33$^{\prime\prime}$ \\
  1.7978377  d & 1.797733 d  & ASASSN-V J161359.09-511123.1 & TIC 409764727 & 16 13 59.09 & -51 11 23.10 &    & \\
  \hline
   ASASSN-V J161358.45-522316.0  &   &  &  &   &   & 1 &    \\
   = TIC 409749495  &   &  &  &   &   &   &    \\
  12.4356049 d & 12.4365535 d& ASASSN-V J161358.53-522317.0 & TIC 409749495 & 16 13 58.53 & -52 23 16.98 &    & 12$^{\prime\prime}$ \\
  0.9461406  d & 0.9461341 d & Gaia DR3 5933200077718865536 & TIC 409749496 & 16 13 57.24 & -52 23 17.09 &    & \\
  \hline
   ASASSN-V J163512.52-360631.4  &   &  &  &   &   &   &    \\
   = TIC 292375578  &   &  &  &   &   &   &    \\
  5.44476510 d & 5.4449524 d & ASASSN-V J163512.54-360631.3 & TIC 292375578 & 16 35 12.54 & -36 06 31.32 &    & 42$^{\prime\prime}$ \\
  0.3932435  d & 0.393351 d  & Gaia DR3 6020312219255157632 & TIC 292375700 & 16 35 12.37 & -36 07 13.66 &    & \\
   \hline
   ASASSN-V J164351.65-451529.1  &   &  &  &   &   & 1 &    \\
   = TIC 234747500  &   &  &  &   &   &   &    \\
  3.53592241 d & 3.5353891 d & ASASSN-V J164351.72-451529.8 & TIC 234747500 & 16 43 51.66 & -45 15 29.59 &    & 5$^{\prime\prime}$ \\
  2.1967167  d & 2.1967355 d & Gaia DR3 5943275349354556800 & TIC 234742498 & 16 43 51.37 & -45 15 28.15 &    & \\
  \hline
   ASASSN-V J172709.26-335241.9  &   &  &  &   &   &   &    \\
   = TIC 160364797  &   &  &  &   &   &   &    \\
  8.57018532 d & 8.5686061 d & ASASSN-V J172709.33-335241.7 & TIC 160364797 & 17 27 09.26 & -33 52 41.74 &    & 44$^{\prime\prime}$ \\
  2.0385762  d & 2.0389 d    & ASASSN-V J172706.11-335255.7 & TIC 160364837 & 17 27 05.95 & -33 52 57.40 &    & \\
  \hline
   ASASSN-V J180536.79-345847.5  &   &  &  &   &   &   &    \\
   = TIC 194787815  &   &  &  &   &   &   &    \\
  4.21447625 d & 4.21445 d   & ASASSN-V J180536.78-345847.7 & TIC 194787815 & 18 05 36.78 & -34 58 47.68 &    & 33$^{\prime\prime}$ \\
  3.9863251  d & 3.986397 d  & OGLE-BLG-ECL-300012          & TIC 194787785 & 18 05 39.20 & -34 59 01.72 &    &  \\
  \hline
   ASASSN-V J180631.44-330249.1  &   &  &  &   &   &   &    \\
   = TIC 55174383  &   &  &  &   &   &   &    \\
  4.82314906 d & 4.8225907 d & OGLE-BLG-ECL-308938          & TIC 55174383 & 18 06 31.50 & -33 02 47.51 &    & 24$^{\prime\prime}$ \\
  0.9195337  d & 0.9196706 d &  V1013 Sgr                   & TIC 55174417 & 18 06 29.88 & -33 02 34.01 &    & \\
  \hline
   ASASSN-V J191542.38+394842.2  &   &  &  &   &   &   &    \\
   = TIC 121866154  &   &  &  &   &   &   &    \\
  9.45584585 d & 9.4555885 d & KID 04737302                 & TIC 121866154 & 19 15 42.40 & +39 48 42.41 &    & 40$^{\prime\prime}$ \\
  9.5089564  d & 9.5240765 d & KID 04737267                 & TIC 121786759 & 19 15 39.67 & +39 48 17.89 &    & \\
  \hline
   ASASSN-V J195933.04+402914.9  &   &  &  &   &   &  1 &    \\
   = TIC 172877901  &   &  &  &   &   &    &    \\
  4.93139848 d & 4.9313644 d & KID 05310435                 & TIC 172877901 & 19 59 32.93 & +40 29 14.39 &    & 29$^{\prime\prime}$ \\
  0.4416729  d & 0.4416691 d &  V2889 Cyg                   & TIC 172877929 & 19 59 30.92 & +40 28 56.75 &    & \\
  \hline \hline
   \end{tabular}}
 {\small Remarks: $1$ - Systems presented in Table 2 of \cite{2023MNRAS.520.2386R}.}
\end{table*}

\begin{table*}
  \caption{False positive systems from Rowan et al. (2023)  %\cite{2023MNRAS.520.2386R}
  together with a correct source identification.}
%  \centering
  \scalebox{0.84}{
  \begin{tabular}{c c c c c c c c}\\[-3mm]
\hline \hline\\[-3mm]
  Information from            &  Published/catalog  &  Identification & TESS number &     RA     &     DE     & Remark  & Angular distance  \\
  \cite{2023MNRAS.520.2386R}  &  period   &              &   &  [J2000.0] & [J2000.0]  &         &   \\
  \hline
   ASASSN-V J195948.25+302430.1  &   &  &  &   &   &   &    \\
   = TIC 103239441  &   &  &  &   &   &   &    \\
  3.62421057 d & 3.6239759 d & ASASSN-V J195948.42+302428.4 & TIC 103239441 & 19 59 48.42 & +30 24 28.37 &    & 62$^{\prime\prime}$ \\
  3.1588364  d & 3.1588172 d &  V1023 Cyg                   & TIC 103239409 & 19 59 53.19 & +30 24 20.48 &    & \\
  \hline
   ASASSN-V J200625.05+194147.7  &   &  &  &   &   &  1 &    \\
   = TIC 282775300  &   &  &  &   &   &    &    \\
  2.17992168 d & 2.1800691 d & WISE J200624.9+194143        & TIC 282775300 & 20 06 24.96 & +19 41 43.91 &    & 12$^{\prime\prime}$ \\
  1.9688380  d & 1.9688193 d & Gaia DR3 1822663831299159936 & TIC 282775321 & 20 06 24.67 & +19 41 55.18 &    & \\
  \hline
   ASASSN-V J202043.12+350856.5  &   &  &  &   &   &   &    \\
   = TIC 135629007  &   &  &  &   &   &   &    \\
  3.71303190 d & 3.7136304 d & ZTF J202042.95+350857.9      & TIC 135629007 & 20 20 42.95 & +35 08 57.95 &    & 36$^{\prime\prime}$ \\
  6.4106845  d & 6.4108 d    & Gaia DR3 2057131318654167040 & TIC 135628948 & 20 20 40.75 & +35 09 21.96 &    & \\
  \hline
   ASASSN-V J205058.15+455201.5  &   &  &  &   &   &   &    \\
   = TIC 356512690  &   &  &  &   &   &   &    \\
  2.27948510 d & 2.2790893 d & ASASSN-V J205058.15+455201.5 & TIC 356512690 & 20 50 58.08 & +45 52 00.90 &    & 30$^{\prime\prime}$ \\
  1.4986448  d & 1.4986394 d & Gaia DR3 2166372815406265344 & TIC 356512678 & 20 51 00.86 & +45 52 06.80 &    &  \\
  \hline
   ASASSN-V J205714.16+362434.7  &   &  &  &   &   &   &    \\
   = TIC 195671787  &   &  &  &   &   &   &    \\
  7.93777260 d & 7.9368943 d & ASASSN-V J205714.18+362434.9 & TIC 195671787 & 20 57 14.18 & +36 24 34.96 &    & 50$^{\prime\prime}$ \\
  0.3394848  d & 0.6789566 d & WISE J205716.5+362515        & TIC 195671829 & 20 57 16.58 & +36 25 15.13 & double period & \\
  \hline
   ASASSN-V J210203.83+413206.9  &   &  &  &   &   &  1 &    \\
   = TIC 314302488  &   &  &  &   &   &    &    \\
  1.91735306 d & 1.9174802 d & ZTF J210203.56+413207.2      & TIC 314302488 & 21 02 03.57 & +41 32 07.29 &    & 30$^{\prime\prime}$ \\
  4.4563486 d  & 8.9096 d    &  V1718 Cyg                   & TIC 314302496 & 21 02 01.10 & +41 31 56.89 & double period & \\
  \hline
   ASASSN-V J215220.51+494311.9  &   &  &  &   &   &   &    \\
   = TIC 384526780  &   &  &  &   &   &   &    \\
  1.76028791 d & 1.7602651 d & WISE J215220.5+494311        & TIC 384526780 & 21 52 20.51 & +49 43 11.86 &    & 33$^{\prime\prime}$ \\
  2.0205318  d & 2.0191008 d & Gaia DR3 1979662856104556672 & TIC 384371717 & 21 52 19.01 & +49 42 42.44 &    & \\
  \hline
   ASASSN-V J215327.68+523450.0  &   &  &  &   &   &   &    \\
   = TIC 430167724  &   &  &  &   &   &   &    \\
  3.39997885 d & 3.3999903 d & ASASSN-V J215327.65+523449.6 & TIC 430167724 & 21 53 27.65 & +52 34 49.58 &    & 31$^{\prime\prime}$ \\
  2.0933696  d & 2.0932828 d & ZTF J215324.41+523500.3      & TIC 430167715 & 21 53 24.42 & +52 35 00.38 &    & \\
  \hline
   ASASSN-V J221421.13+525718.1  &   &  &  &   &   &   &    \\
   = TIC 430678543  &   &  &  &   &   &   &    \\
  4.34214829 d & 4.3421799 d & ASASSN-V J221420.91+525718.6 & TIC 430678543 & 22 14 20.91 & +52 57 18.65 &    & 55$^{\prime\prime}$ \\
  1.5747661  d & 0.7869864 d & Gaia DR3 2004340398159008768 & TIC 430678597 & 22 14 18.59 & +52 56 27.46 & half period & \\
  \hline
   ASASSN-V J221443.29+612611.8  &   &  &  &   &   & 1 &    \\
   = TIC 336098020  &   &  &  &   &   &   &    \\
  19.29902233 d& 19.2949942 d& ASASSN-V J221443.31+612611.9 & TIC 336098020 & 22 14 43.31 & +61 26 11.94 &    & 79$^{\prime\prime}$ \\
  2.5506271   d& 2.5504824 d & ASASSN-V J221453.49+612643.2 & TIC 336097969 & 22 14 53.49 & +61 26 43.15 &    & \\
  \hline
   ASASSN-V J223355.43+562314.9  &   &  &  &   &   &   &    \\
   = TIC 64558868  &   &  &  &   &   &   &    \\
  2.60096408 d & 5.2012926 d & ASASSN-V J223355.45+562315.3 & TIC 64558868 & 22 33 55.45 & +56 23 15.36 & double period & 32$^{\prime\prime}$ \\
  0.4710463  d & 0.4710574 d & WISE J223354.3+562346        & TIC 64558820 & 22 33 54.40 & +56 23 46.28 &     & \\
  \hline
   ASASSN-V J230746.60+520824.6  &   &  &   &   &   &   &    \\
   = TIC 252683086  &   &  &   &   &   &   &    \\
  3.22349972 d & 3.2234997 d & ASASSN-V J230746.60+520824.6 & TIC 252683086 & 23 07 46.58 & +52 08 24.84 &     & 26$^{\prime\prime}$ \\
  7.2878876  d & 7.2898    d & Gaia DR3 1995202460088899840 & TIC 252683096 & 23 07 49.33 & +52 08 38.85 &     & \\
   \hline \hline
   \end{tabular}}
 {\small Remarks: $1$ - Systems presented in Table 2 of \cite{2023MNRAS.520.2386R}.}
\end{table*}

\section{Conclusions}  \label{Conclusions}

Quadruple eclipsing stellar systems of 2+2 nature, or triple stars with two eclipsing periods of
2+1 architecture, bring unique insight into the multiple-star dynamics and evolution. Hence, their
detection is highly valuable, especially in large photometric databases. However, such a detection
should be done rigorously, and with all available means. This is especially true when using TESS
data due to the relatively low angular resolution per-pixel. Therefore, careful identification of
false positives is a critical step in any efforts aimed at increasing the number of genuine
multiply-eclipsing multiple stellar systems.

\section*{Acknowledgments}
An anonymous referee is greatly acknowledged for his/her helpful suggestions, greatly improving
the manuscript. The author thank the ASAS-SN, ZTF, Atlas, ASAS, and TESS teams for making all of
the observations easily public available. The research of P.Z. was also supported by the project
{\sc Cooperatio - Physics} of Charles University in Prague. This research has made use of the
SIMBAD and VIZIER databases, operated at CDS, Strasbourg, France and of NASA Astrophysics Data
System Bibliographic Services. This work has made use of data from the European Space Agency (ESA)
mission {\it Gaia} (\url{https://www.cosmos.esa.int/gaia}), processed by the {\it Gaia} Data
Processing and Analysis Consortium (DPAC,
\url{https://www.cosmos.esa.int/web/gaia/dpac/consortium}). Funding for the DPAC has been provided
by national institutions, in particular the institutions participating in the {\it Gaia}
Multilateral Agreement.

\section*{Data availability}

All the data used in the manuscript, which are not already included in the Tables, will be shared
on reasonable request to the corresponding author.


\begin{thebibliography}{}
 \bibitem[\protect\citeauthoryear{{\'A}d{\'a}m et al.}{2023}]{2023A&A...674A.170A} {\'A}d{\'a}m R.~Z., Hajdu T., B{\'o}di A., Hajdu R., Szklen{\'a}r T., Moln{\'a}r L., 2023, A\&A, 674, A170. doi:10.1051/0004-6361/202346006
 \bibitem[\protect\citeauthoryear{Gaia Collaboration et al.}{2023}]{2023A&A...674A...1G} Gaia Collaboration, Vallenari A., Brown A.~G.~A., Prusti T., de Bruijne J.~H.~J., Arenou F., Babusiaux C., et al., 2023, A\&A, 674, A1. doi:10.1051/0004-6361/202243940
 \bibitem[\protect\citeauthoryear{Heinze et al.}{2018}]{2018AJ....156..241H} Heinze A.~N., Tonry J.~L., Denneau L., Flewelling H., Stalder B., Rest A., Smith K.~W., et al., 2018, AJ, 156, 241. doi:10.3847/1538-3881/aae47f
 \bibitem[\protect\citeauthoryear{Kochanek et al.}{2017}]{2017PASP..129j4502K} Kochanek C.~S., Shappee B.~J., Stanek K.~Z., Holoien T.~W.-S., Thompson T.~A., Prieto J.~L., Dong S., et al., 2017, PASP, 129, 104502. doi:10.1088/1538-3873/aa80d9
 \bibitem[\protect\citeauthoryear{Masci et al.}{2019}]{2019PASP..131a8003M} Masci F.~J., Laher R.~R., Rusholme B., Shupe D.~L., Groom S., Surace J., Jackson E., et al., 2019, PASP, 131, 018003. doi:10.1088/1538-3873/aae8ac
 \bibitem[\protect\citeauthoryear{Pojmanski}{1997}]{1997AcA....47..467P} Pojmanski G., 1997, AcA, 47, 467. doi:10.48550/arXiv.astro-ph/9712146
 \bibitem[\protect\citeauthoryear{Ricker et al.}{2015}]{2015JATIS...1a4003R} Ricker G.~R., Winn J.~N., Vanderspek R., Latham D.~W., Bakos G. {\'A}., Bean J.~L., Berta-Thompson Z.~K., et al., 2015, JATIS, 1, 014003. doi:10.1117/1.JATIS.1.1.014003
 \bibitem[\protect\citeauthoryear{Rowan et al.}{2023}]{2023MNRAS.520.2386R} Rowan D.~M., Jayasinghe T., Stanek K.~Z., Kochanek C.~S., Thompson T.~A., Shappee B.~J., Holoien T.~W.-S., et al., 2023, MNRAS, 520, 2386. doi:10.1093/mnras/stad021
 \bibitem[\protect\citeauthoryear{Shappee et al.}{2014}]{2014ApJ...788...48S} Shappee B.~J., Prieto J.~L., Grupe D., Kochanek C.~S., Stanek K.~Z., De Rosa G., Mathur S., et al., 2014, ApJ, 788, 48. doi:10.1088/0004-637X/788/1/48
 \bibitem[\protect\citeauthoryear{Watson, Henden, \& Price}{2006}]{2006SASS...25...47W} Watson C.~L., Henden A.~A., Price A., 2006, SASS, 25, 47
 \bibitem[\protect\citeauthoryear{Zasche}{2024}]{2024A&A...688A..41Z} Zasche P., 2024, A\&A, 688, A41. doi:10.1051/0004-6361/202450463
 \end{thebibliography}
\end{document}